\documentclass[aps,prl,twocolumn,superscriptaddress,showpacs,floatfix]{revtex4-1}
\usepackage{makecell}
\usepackage{graphicx}
\usepackage{dcolumn}
\usepackage{bm}
\usepackage{times}
\usepackage{color}
\usepackage{hyperref}
\usepackage{graphicx}
 \usepackage{epstopdf}

\usepackage{textcomp}
\begin{document}

\title{Extremely large magnetoresistance and chiral anomaly in the nodal-line semimetal ZrAs$_2$}

\author{Junjian Mi}
\affiliation{School of Physics, Zhejiang University, Hangzhou,
310058, China } \affiliation{State Key Laboratory of Silicon and
Advanced Semiconductor Materials, Zhejiang University, Hangzhou,
310027, China}
\author{Sheng Xu}
\affiliation{School of Physics, Zhejiang University, Hangzhou,
310058, China }

\author{Shuxiang Li}
\affiliation{School of Physics, Zhejiang University, Hangzhou,
310058, China } \affiliation{State Key Laboratory of Silicon and
Advanced Semiconductor Materials, Zhejiang University, Hangzhou,
310027, China}
\author{Chenxi Jiang}
\affiliation{School of Physics, Zhejiang University, Hangzhou,
310058, China } \affiliation{State Key Laboratory of Silicon and
Advanced Semiconductor Materials, Zhejiang University, Hangzhou,
310027, China}
\author{Zheng Li}
\affiliation{School of Physics, Zhejiang University, Hangzhou,
310058, China } \affiliation{State Key Laboratory of Silicon and
Advanced Semiconductor Materials, Zhejiang University, Hangzhou,
310027, China}
\author{Qian Tao}
\affiliation{School of Physics, Zhejiang University, Hangzhou,
310058, China }
\author{Zhu-An Xu}\email{zhuan@zju.edu.cn}
\affiliation{School of Physics, Zhejiang University, Hangzhou,
310058, China } \affiliation{State Key Laboratory of Silicon and
Advanced Semiconductor Materials, Zhejiang University, Hangzhou,
310027, China} \affiliation{Hefei National Laboratory, Hefei,
230088, China} \affiliation{ Collaborative Innovation Centre of
Advanced Microstructures, Nanjing University, Nanjing, 210093,
China}


\date{\today}

\begin{abstract}
We performed the detailed magnetotransport measurements and first
principle calculations to study the electronic properties of the
transition metal dipnictides ZrAs$_2$, which is a topological
nodal-line semimetal. Extremely large unsaturated
magnetoresistance (MR) which is up to $1.9\times10^{4}$ \% at 2 K
and 14 T was observed with magnetic field along the $c$-axis. The
nonlinear magnetic field dependence of Hall resistivity indicates
the multi-band features, and the electron and hole are nearly
compensated according to the analysis of the two-band model, which
may account for the extremely large unsaturated MR at low
temperatures. The evident Shubnikov-de Haas (SdH) oscillations at
low temperatures are observed and four distinct oscillation
frequencies are extracted. The first principle calculations and
angle-dependent SdH oscillations reveal that the Fermi surface
consists of three pockets with different anisotropy. The observed
twofold symmetry MR with electric field along the $b$-axis
direction is consistent with our calculated Fermi surface
structures. Furthermore, the negative magnetoresistance (NMR) with
magnetic field in parallel with electric field is observed,
which is an  evident feature of the chiral anomaly.

\end{abstract}
\maketitle
\setlength{\parindent}{1em}
\section{Introduction}
The study of topological semimetals have attracted tremendous interests due to their novel electronic structure and unconventional transport properties in condensed-matter physics in recent years \cite{ref1,ref2,ref3},
 such as large longitudinal magnetoresistance, high mobility, nontrivial Berry phase and chiral-anomaly induced negative longitudinal MR
\cite{ref4,ref5,ref6,ref7,ref8,ref9,ref10}.
In these materials, the topological semimetallic states are
characterized by band touching points or lines between conduction
and valance bands in momentum space. Dirac and Weyl semimetals are
characterized by band crossings at a single point in $k$-space,
exhibiting linear dispersion and featuring fourfold and twofold
degeneracies, respectively. Dirac points could be divided into two
Weyl points by breaking time-reversal symmetry or inversion
symmetry. Dirac semimetals have been confirmed in Cd$_3$As$_2$ and
Na$_3$Bi \cite{ref11,ref12,ref13,ref14} and Weyl semimetals have
been theoretically proposed and experimentally observed in MX
(M=Ta or Nb; X=P or As), WTe$_2$ and (Nb,Ta)IrTe$_4$
\cite{ref15,ref16,ref17,ref18,ref19,ref20,ref21}.
In contrast to Dirac or Weyl semimetals, topological nodal line
semimetals showcase a unique characteristic: one-dimensional
trajectory within the $k$-space, forming either an open path or a
closed loop. Several candidate materials have been identified
within the realm of nodal line semimetals, including PbTaSe$_2$,
PtSn$_4$, ZrSiX (X = S, Se, Te), and the CaP$_3$ family (A = Ca,
Ba, Sr; B = P, As). Each of these materials exhibits intriguing
topological properties \cite{ref22,ref23,ref24,ref25,ref26,ref27}.

The transition metal dipnictides (TMDPs) with the formula MX$_2$
(where M = Nb or Ta; X = As or Sb) exhibit a crystalline structure
characterized by the symmetry of \textit{C2/m} (No. 12), and they
have attracted considerable attention primarily owing to their
possession of compelling properties, including extremely large
magnetoresistance (XMR) and negative magnetoresistance (NMR)
\cite{ref9,ref10,ref28}. Recently, TMDPs ZrX$_2$ (where X = P or
As) with the space group \textit{Pnma} (No. 62) were theoretically
predicted to be topological nodal-line semimetals, garnering
increasing attention. An extremely large magnetoresistance in
ZrP$_2$ has been detected, which might be induced by the
compensation of electrons and holes. The compensation of electrons
pockets and hole pockets were further confirmed by angle-resolved
photoemission spectroscopy (ARPES) as well as the first-principle
calculations \cite{ref29,ref30}. The band structure of ZrP$_2$ is
characterized by a nodal loop in the $k_x$ = 0 plane without
considering spin-orbit coupling (SOC) and a small gap along the
nodal loop will open when considering SOC according to the
first-principle calculations \cite{ref29}. Compared with ZrP$_2$,
the gap of ZrAs$_2$ along the node line should be larger due to
the enhanced spin-orbit coupling\cite{ref6}.

Quantum oscillations, Hall resistivity, and anisotropic MR in ZrAs$_2$ have recently been reported
\cite{ref31}. However, the topological properties have yet to be
confirmed and the transport properties are worth revisiting with
enhanced sample quality. Only one oscillation frequency (i.e.,
single pocket in Fermi surface) was observed from the SdH oscillations with the current applying along the
$b$- axis and the magnetic field in the $ac$ plane of ZrAs$_2$. In
contrast, multiple pockets were detected in de Haas-van Alphen
(dHvA) oscillations. Furthermore, recent ARPES experiments have
reported the surface bands and bulk states of ZrAs$_2$, suggesting
multi-band feature \cite{ref32}. Such a discrepancy between
transport and magnetic measurements could result from the low
sample quality indicated by the lower residual resistivity ratio (RRR)
of the previous samples grown by the flux method.

In this work, we reported the detailed magnetotransport properties
and electronic structure of the topological nodal-line semimetal
ZrAs$_2$. We used chemical vapor transport method to grow high
quality crystals of ZrAs$_2$ to investigate topologital
properties. The Hall resitivity and longitudinal MR measurements
reveal the concentration of electron ($n_e\approx 3.58\times10^{20}
cm^{-3}$) with a mobility of $\mu_e$ $\approx$ 5600 $cm^{2} V^{-1}s^{-1}$
and the concentration of hole ($n_h \approx 3.47\times10^{20} cm^{-3}$)
with a mobility of $\mu_h$ $\approx$ 5164 $cm^{2} V^{-1}s^{-1}$ at 2 K,
according to the analysis of the two-band model. The XMR up to
$1.9\times10^{4}$ \% is oberved at 2 K and 14 T, which is
considered to stem from the compensation mechanism \cite{ref29}.
The SdH oscillations  at low temperature and high magnetic field
are observed and four frequencies  F$_1 =$ 155 T, F$_2 =$ 178 T,
F$_3 =$ 443 T,  F$_4 =$ 579 T  are extracted.  The nontrivial
Berry phases are obtained from the plot of the Landau level index
fan diagram. Moreover, we observed anistropic MR with two-fold
symmetry when the current is applied along the $b$ axis and the
magnetic field lies within the $ac$ plane, consistent with the
two-fold Fermi surfaces (FSs) structure predicted by the DFT
calculations. Furthermore, we observed the evident NMR when the
electric field is in parallel with magnetic field applying along
the $b$ axis. According to our analysis and previous theoretical
works\cite{ref33}, we argued that the NMR should be induced by the chiral
anomaly. Our study has provided compelling evidence for the
topological properties of nodal-line semimetals in ZrAs$_2$.

\section{EXPERIMENTAL METHODS}

Single crystals of ZrAs$_2$ were synthesized by the chemical vapor transport method. The Zr power and As power were weighted with a molar ratio of 1:2 in sealed fused-silica tubes with 4 mg/mL grain of I$_2$ added to promote crystal growth. The quartz tubes were heated to 850 $^{\circ}$C over 2 days, kept at 850 $^{\circ}$C  for 1 week and cooled to room temperature over 1 day, as described in Ref \cite{ref34}. Then needle-like crystals were obtained.
The atomic composition of the ZrAs$_2$ single crystal was determined using energy dispersive X-ray spectroscopy (EDS), revealing a Zr to As ratio of 1:1.92. This observed ratio aligns closely with the nominal ratio of 1:2, taking into account experimental errors. A suitable crystal was selected and mounted on a Bruker D8 Venture diffractometer to measure the single crystal X-ray diffraction (XRD). The crystal was kept at 170.0 K during data collection. Using Olex2 \cite{ref35}, the structure was solved with the SHELXT \cite{ref36} structure solution program using Intrinsic Phasing and refined with the SHELXL \cite{ref37} refinement package using Least Squares minimisation.

\begin{figure}
\centering
  \includegraphics[width=0.5\textwidth]{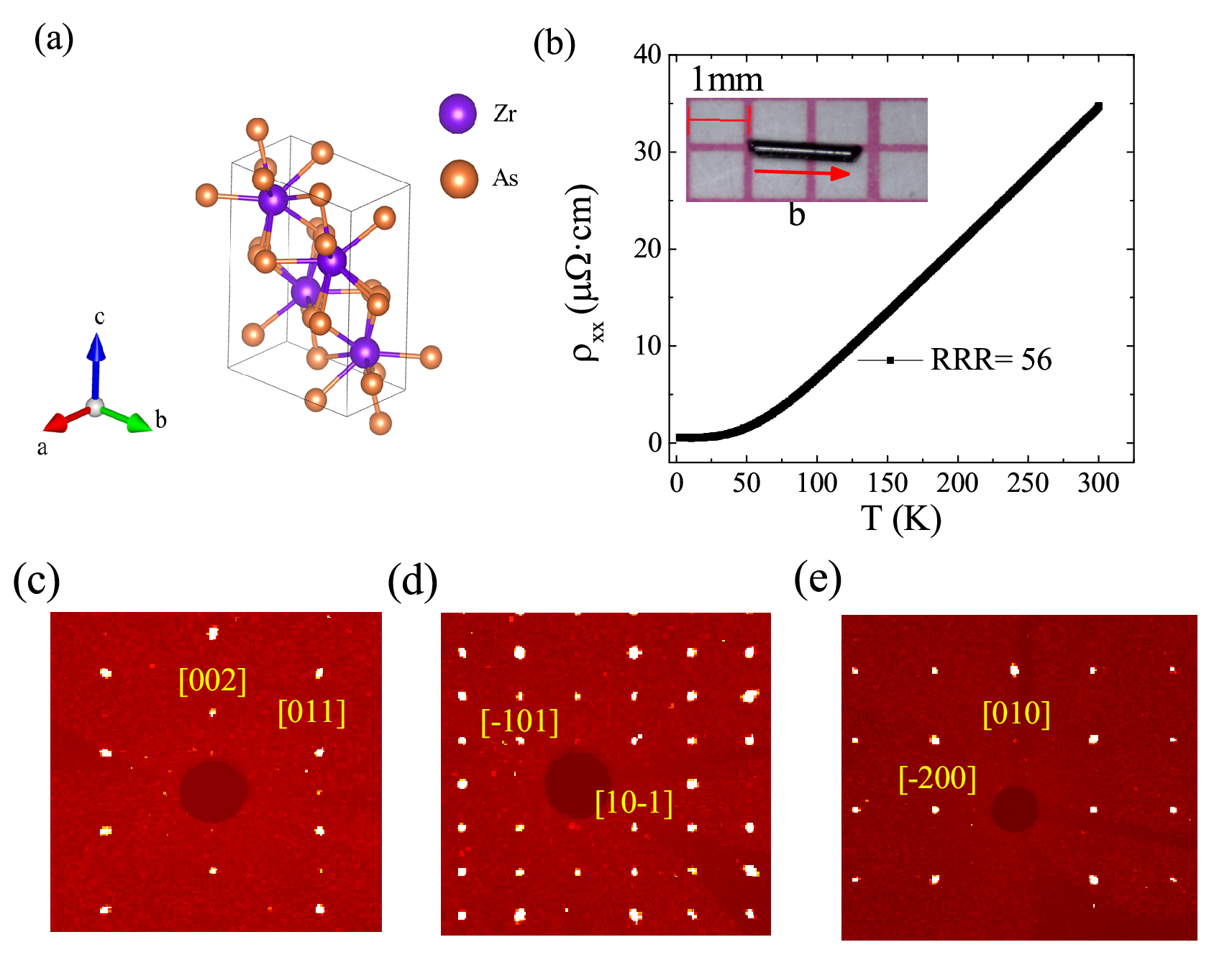}\\
  \caption{(a) Crystaline structure of ZrAs$_2$. (b) Temeperature-dependent resistivity of the single crystal ZrAs$_2$. The inset shows a photograph of the grown ZrAs$_2$ single crystal. (c)-(e) Single-crystal XRD on ZrAs$_2$ with Diffracted reflections in ( $0kl$ ) plane, ( $h0l$ ) plane ( $hk0$ ) plane, respectively.
 }\label{1}
\end{figure}

The measurements of resistivity and magnetic transport properties were performed on a Quantum Design physical property measurement system (PPMS-DynaCool 14 T). The DFT calculations of electronic structure of bulk ZrAs$_2$ were performed by using the Vienna ab initio simulation package (VASP)
\cite{ref38}
, with the generalized gradient approximation (GGA) in the Perdew-Burke-Ernzerhof type as the exchange-correlation energy \cite{ref39}.
 The cutoff energy was set to 350 eV and $\Gamma$-centered 8 $\times$ 15 $\times$ 6 mesh were sampled over the Brillouin zone integration. The lattice constant $a$ $=$ 6.7964 {\AA}, $b$ $=$ 3.6848 {\AA}, $c$ $=$ 9.0214 {\AA} was used for all the calculations. The tight-binding model of ZrAs$_2$ was constructed by the Wannier90 with 4$d$ orbitals of Zr and 4$p$ orbitals of As, which based on the maximally-localized Wannier functions
\cite{ref40}
\cite{ref41}.

\section{Results and Discussions}

The crystal structure of ZrAs$_2$ is presented in Fig. 1(a) with
the space group of $Pnma$ (No. 62). Fig. 1(b) shows the
temperature-dependent $\rho_{xx}$ exhibiting a typical metallic
behavior with the electric current along the $b$-axis. The
residual resistivity $\rho_{xx}$( 2K) is 0.53  $\mu\Omega\cdot$cm at 2 K and residual resistance ratio
(RRR=$\rho_{xx}$(300 K)/$\rho_{xx}$(2 K)) is 56 which is higher than
previously reported value in Ref. \cite{ref31}. A higher $RRR$ indicates a higher quality of single crystals. The inset shows a
picture of the needle shaped single crystal ZrAs$_2$ with the easy
growth direction along the $b$-axis. Figs. 1(c-e)  presents  the
diffracted reflections in ( $0 k l$ ) plane, ( $h 0 l$ ) plane and ( $h k 0$ ) plane, respectively. The lattice constants of ZrAs$_2$ were
obtained as $a$ $=$ 6.7964(3) {\AA}, $b$ $=$ 3.6848(2) {\AA}, $c$ $=$
9.0214(4) {\AA}, which are consistent with the previous reported
results \cite{ref34}.

\begin{figure}[htbp]
\centering
  \includegraphics[width=0.5\textwidth]{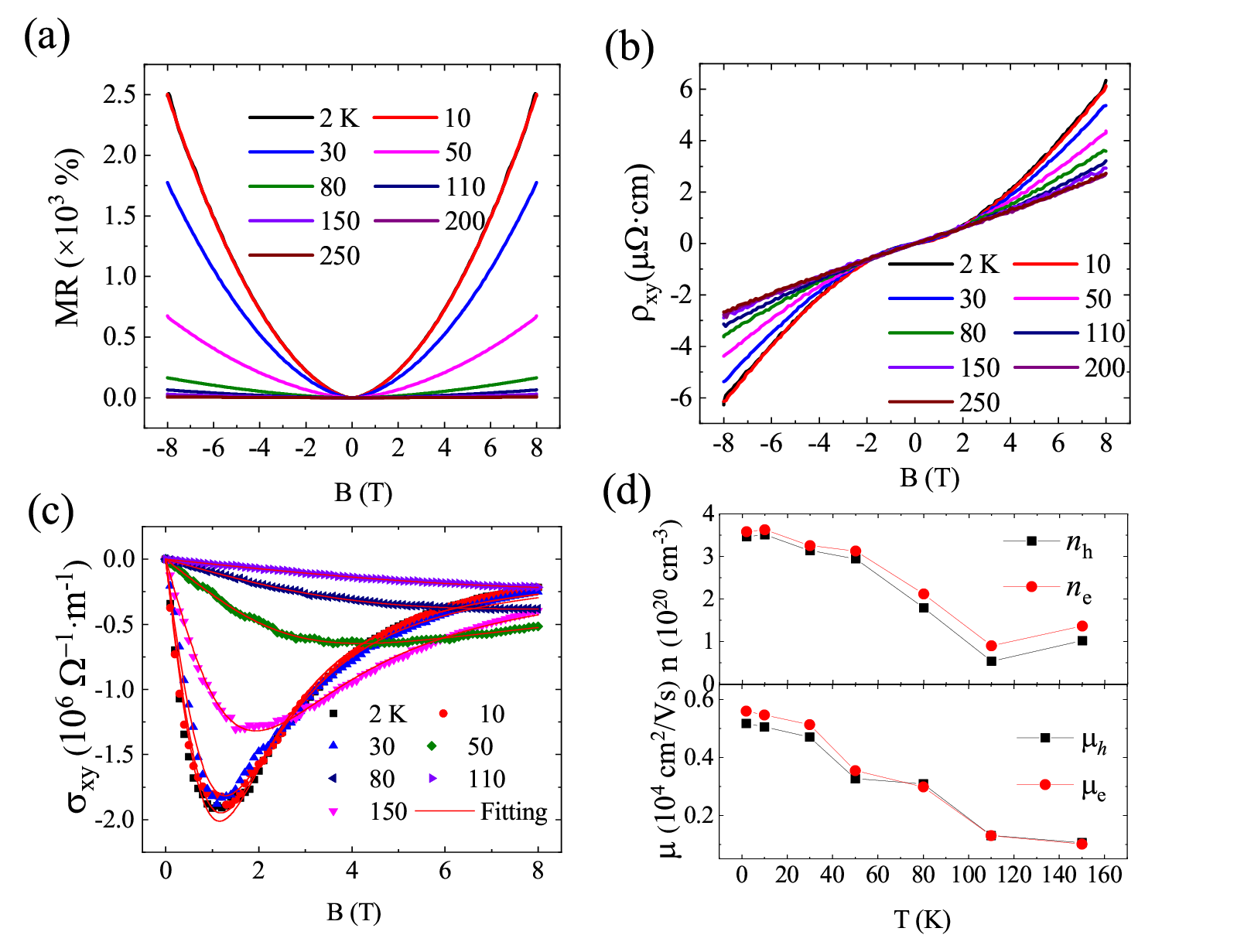}\\
\caption{
(a) The longitudidual MR with the currrent along the $b$ axis and magnetic field under the $ac$ plane at various temperatures. (b) Hall resistivities versus magnetic field at different temperatures. (c) $\sigma_{xy}$ versus magnetic field at different temperatures. The red curves are the fitting results by two-band model. (d) The temperature dependence of carrier densities and mobilities of the electrons and holes. } \label{2}
\end{figure}

Fig. 2(a) shows the field-dependent longitudinal MR of ZrAs$_2$ at
various temperatures, where MR is defined as MR =
$(\rho_{xx}(B)-\rho_{xx}(0))/(\rho_{xx}(0))\times 100\%$. To
eliminate the Hall contribution, the $\rho_{xx}(B)$ is calculated
from $\rho_{xx}(B)=(\rho_{xx}(-B)+\rho_{xx}(B))/2$ . An extremely
large unsaturated MR of up to $2.5\times10^{3}$ \% is observed at
2 K and 8 T. The MR decreases as temperature increases,
reaching $1.6\times10^{2}$\% at 80 K and 8 T. To explore the
origin of this large and unsaturated MR, we also measured the Hall
resistivity. The measurements of $\rho_{xy}(B)$ at different
temperatures provide further transport features, as shown in Fig.
2(b). The Hall resistivity exhibits nonlinear magnetic field
dependence at low temperatures, implying its multi-band
properties. Furthermore, the mobility and concentration of
electron and hole can be obtained by a two-band model fitting. Firstly, we
calculated the $\sigma_{xy}$ according to the  formula $
\sigma_{xy}=-\rho_{xy}/((\rho_{xy})^2+(\rho_{xx})^2) $, then we
 fit the $\sigma_{xy}$ using the two-band model formula as the following.
\begin{equation}\label{equ3}
\sigma_{xy}(B) =(\frac{n_h \mu_h^2}{1+(\mu_h B)^2}-\frac{n_e \mu_e^2}{1+(\mu_e B)^2})eB
\end{equation}
 where $n_{e,h}$ and $\mu_{e,h}$ represent the concentration and mobility of electrons and holes, respectively. As shown in Fig. 2(c), the fitted red curves are consistent with the experimental data, from which the temperature-dependent concentrations and mobilities are extracted and plotted in Fig. 2(d). At 2 K, the concentration of electron $n_e\approx 3.58\times10^{20}$  $cm^{-3}$ which is almost equal to the concentration of hole $n_h\approx 3.47\times10^{20}$ $cm^{-3}$. The mobilities increase with decreasing temperature. $\mu_e $ and $\mu_h $ reach up to 5600 $cm^{2}V^{-1}s^{-1}$ and 5164 $cm^{2}V^{-1}s^{-1}$ at 2 K, respectively. The high mobilities further illustrate the good quality of single crystals. According to these results and analysis, the compensation mechanism of electrons and holes can account for the origin of the XMR at low temperature. Similarly, the compensation mechanism of electrons and holes induced XMR has been reported in other topological semimetals such as TaAs$_2$, NbAs$_2$ and ZrP$_2$ \cite{ref9,ref10,ref29}.


 Quantum oscillation experiments serve as a potent approach to investigate the properties of Fermi surfaces \cite{ref42,ref43}. Fig. 3(a) shows the resistivity as a function of magnetic field at low temperatures with the electric current along the $b$-axis and the magnetic field  along the $c$-axis. The SdH oscillations can be clearly observed at low temperatures and high magnetic fields. The inset shows obvious oscillations around 14 T. The oscillations gradually decreases as temperature increases and almost disappears at 12 K. The oscillatory components of the SdH oscillations versus 1/$B$ can be described by the Lifshitz-Kosevich (LK) formula \cite{ref44,ref45,ref46,ref47}:
\begin{equation}\label{equ1}
\Delta \rho \propto \frac{\lambda T}{cosh(\lambda T)}e^{-\lambda T_D}sin[2\pi(\frac{F}{B}-\frac{1}{2}+\beta+\delta)]
\end{equation}
where $\lambda=(2\pi^2 k{_B} m^*)/(\hbar eB)$. $T{_D}$ is the Dingle
temperature. The value of $\delta$ depends on the dimensionality,
$\delta=0$ for 2D systems and $\delta=\pm1/8$ for 3D systems.
 $\beta=\phi_B/2\pi$ and $\phi_B$ is the Berry phase.
After substracting a background, the oscillation component $\Delta \rho$ versus 1/$B$ is shown in Fig. 3(b). The oscillation intensity decreases with increasing temperature. Four oscillation frequencies can be identified from the Fast Fourier transform (FFT) analysis with F$_1 =$155 T , F$_2 =$178 T,  F$_3 =$443 T and F$_4 =$579 T  at low temperatures as shown in Fig. 2(c). Furthermore, the harmonic frequency 2 F$_3 =$886 T and 2 F$_4 =$1158 T can also be observed.
\begin{figure}[htbp]
\centering
 \includegraphics[width=0.5\textwidth]{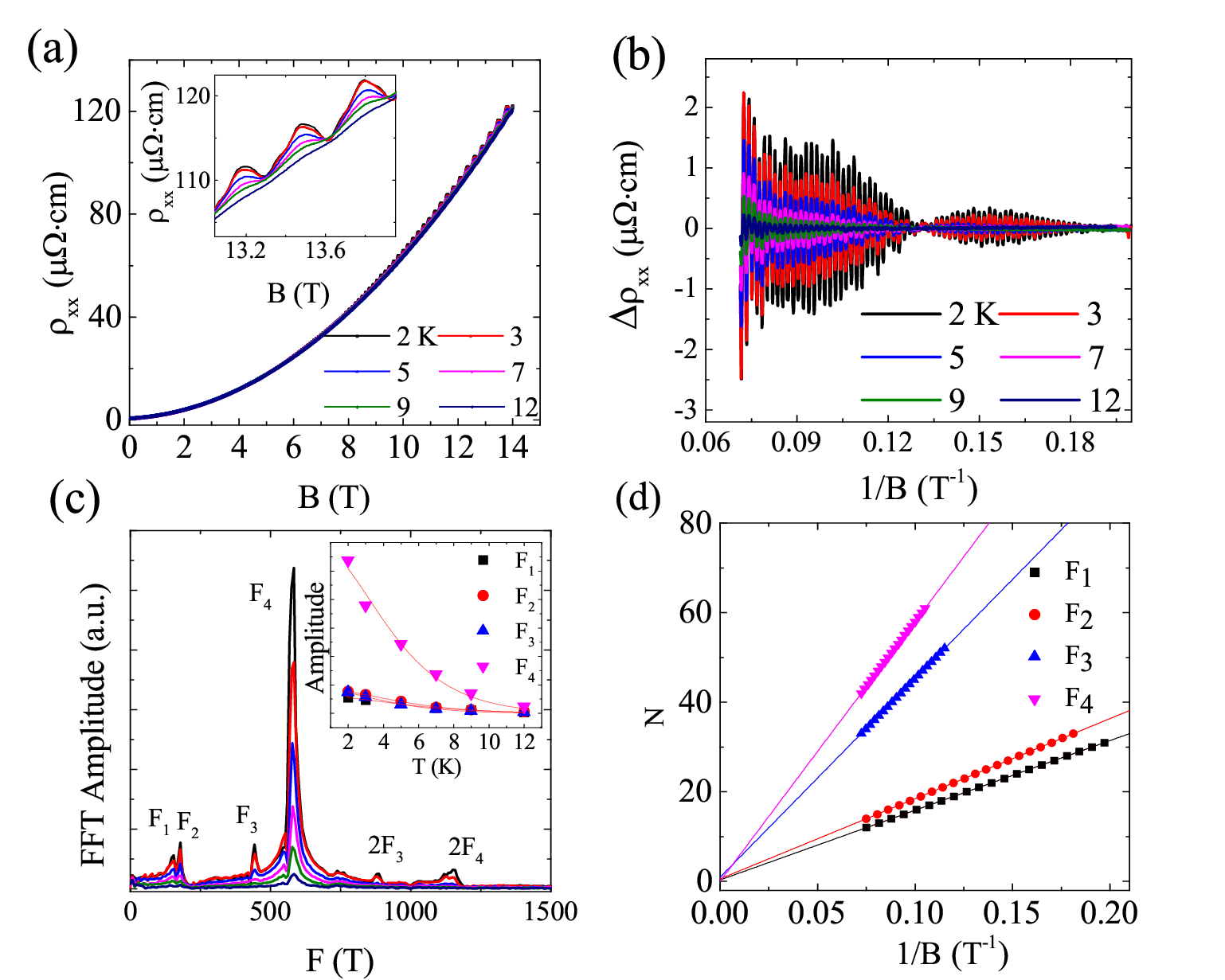}\\
\caption{(a) SdH oscillations at low temperatures with current along the $b$-axis and magnetic field along the $c$-aixs. Inset: enlarged plot of SdH oscillations at high fields. (b)  Oscillatory part of the resistivity $\Delta$$\rho$ as a function of 1/ $B$ at different temperatures. (c) FFT curves indicate the presence of four distinct intrinsic frequencies. Inset shows the temperature dependence of relative FFT amplitude of the frequencies as a function of temperature and the read curves are the fitted by the thermal factor of Lifshitz-Kosevich formula. (d)  Landau-level indices fan diagram for different frequencies as a function of 1/ $B$.}\label{3}
\end{figure}
According to the Onsager relation $F=(\phi{_0}/2\pi^2)=(\hbar/2\pi e)A$$_F$, the oscillation frequency is proportional to the extreme cross-section ($A_F$) of FS normal to the magnetic field. The calculated $A_F$ associated with different frequency were shown in Table 1. The inset of Fig. 3(c) shows the temperature-dependent FFT amplitudes and the red curve are fitted by the thermal factor $R_T=(\lambda T)/sinh(\lambda T)$. The fitting curves are consistent with experiment datas and the effective masses are estimated to be $m^*_1=0.10m_e$, $m^*_2=0.11m_e$ , $m^*_3=0.13m_e$ and $m^*_4=0.12m_e$. The detailed fitting results are listed in Table 1. Compared to previous reports \cite{ref31}, we obtained lighter effective masses, which are coresponding to the Dirac fermions with linear dispersion. This suggests that the electronic properties of this material are more akin to ideal Dirac fermions. Besides, according to the Lifshitz - Onsager quantization rule $N=A_F(\hbar/2\pi eB)-1/2+\beta+\delta$, the  Berry phase can be extracted from the intercept of the linear extrapolation. The Landau indices $N$ as a function of 1/$B$ and the fitting curves are shown in Fig. 3(d). The values of $\phi_B$ from the different $\delta$ are listed in Table 1. The Berry phases of $F_3$ with ($\delta = 1/8$) is 1.1 $\pi$ and  $F_4$ with ($\delta = -1/8$) is 0.8 $\pi$, indicating the nontrivial $\pi$ Berry phases. The nontrivial Berry phases indicate the topological properities of corresponding bands and FSs.


To further investigate the FS topology and compare it with results of the SdH experiments, we performed the DFT calculations. 
The FSs structures in the first Brillouim zone are displayed in
Fig. 4(a)-4(c). ZrAs$_2$ has small $\alpha$ hole pockets
(blue) near U point and  two big $\beta$ hole pockets
as shown in Fig. 4(c) and Fig. 4(d) and three type $\delta$ electron
pockets (purple) near $\Gamma$ point can be devided as $\delta_1$
and $\delta_2$, which is similar to its sister compound ZrP$_2$
\cite{ref6}. In addition, there is also a small electron-like FS
inside $\delta_2$. There are two
additional small Fermi surfaces located inside the $\delta_2$ Fermi surfaces. Moreover, both large Fermi surfaces,
$\beta$  and $\delta_2$, exhibit two-fold symmetry in the first
Brillouin zone.


\begin{table*}
  \centering
\caption{The parameters extracted from SdH oscillations with \emph{B}$\bot$ $c$-axis configuration. The $F_1$, $F_2$, $F_3$ and $F_4$ are the FSs from SdH oscillations; $m^*$/$m_e$ is the ratio of the effective mass to the electron mass; $\phi$$_B$ is the Berry phase.}
  \label{oscillations}
  \setlength{\tabcolsep}{4mm}{
  \begin{tabular}{ccccccccccccc}
  \hline
  \hline

  &Pocket &$Frequency (T)$  &$A_F$(\AA$^{-2}$) & $\phi_B(\delta=-1/8)$ & $\phi_B(\delta=1/8)$ &  $m^*$/$m_e$  &  \\

\hline
  &  $F_1$   &155           &0.015       &1.4$\pi$    & 0.9$\pi$     &   0.10        &\\
  &  $F_2$    &178           &0.017      &1.9$\pi$    & 0.4$\pi$     &   0.11       &\\
  &  $F_3$   &443          &0.042       & 0.6$\pi$    & 1.1$\pi$     &   0.13         &\\
   &  $F_4$  &579          &0.055      & 0.8$\pi$    & 1.3$\pi$     &   0.12         &\\

  \hline
  \hline
  \end{tabular}}
  \end{table*}


Angular-dependent magnetotransport measurements provide a valuable
approach for investigating the detailed structure of the FS
\cite{ref43}. Fig. 4(d) shows the angle-dependent MR as a function
of magnetic field at 2 K. The inset of Fig. 4(d) presents the
definition of $\varphi$ where the magnetic field is always
perpendicular to the current. The evident SdH oscillations are
observed when $H$ rotates from the $a$ axis to $c$ axis.  The
angle-dependent  $\Delta$$\rho$ as a function of 1/ $B$ at different angles are shown in Fig. 4(e). The
itensities of oscillations increase with decreasing $\varphi$
 which is consistent with the tendency of MR.  Fig. 4(f) shows the
angle-dependent frequency of SdH oscillations as the rotation
angle changes from the $a$ axis ($\varphi$ = $0^o$) to $c$ axis
($\varphi$ = $90^o$). As we mentioned hereinbefore, the four
frequencies, $F_1$, $F_2$, $F_3$ and $F_4$, appear as $B // c$. Due
to the complicated FSs, we draw a red dash line in Fig. 4(f) to
carefully tracke the position of these frequencies at different
angles.
\begin{figure}[htbp]
\centering
  \includegraphics[width=0.5\textwidth]{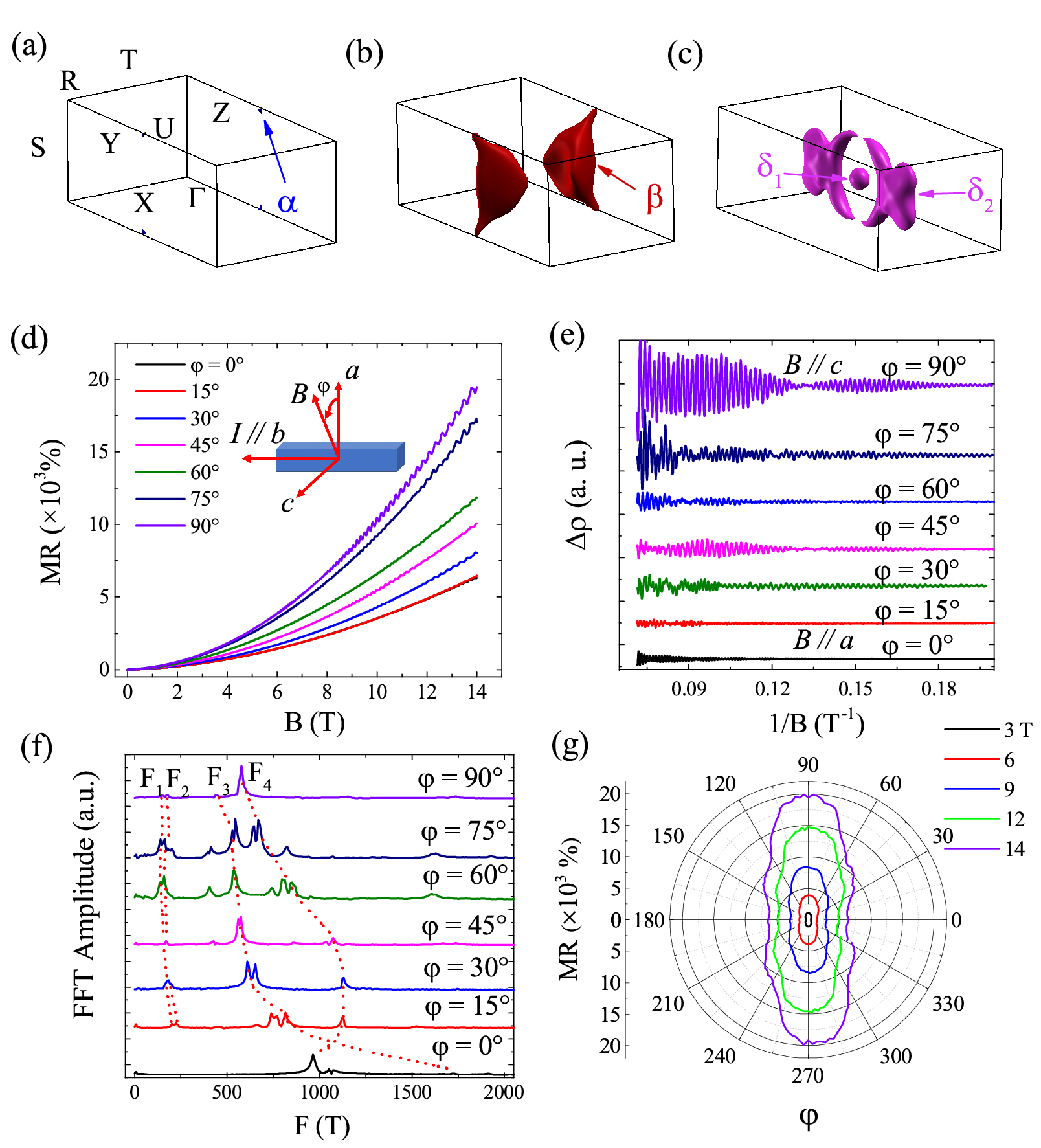}\\
\caption{(a) Brillouin zone and FSs of electron pockets (blue ones). (b) FSs of hole pockets (red ones). (c) FSs of electron pockets (purple ones). (d) Angle-dependent MR as a function of magnetic field of ZrAs$_2$ by titling the magnetic fields from $B$ // $a$ direction to the $B$ // $c$ direction with the current along the  $b$ direction at 2 K. (e) Angle-dependent $\Delta$$\rho$ as a function of 1/ $B$ at different angles. (f) Angle-dependent frequency from SdH oscillations at different angles. (g) Angle-dependent MR at different magnetic field by titling the magnetic fields under the ($ac$) plane.}\label{5}

\end{figure}
The frequencies of $F_1$ and $F_2$ change slightly. However, the
signals of $F_1$ and $F_2$ disappear at $\varphi$ = $0^o$, which
possibly is obscured by the signal from a larger pocket. The
frequencies of $F_3$ increased with decreasing $\varphi$.
However, the frequency of $F_4$ increases first and then decreases
with decreasing $\varphi$. The Fermi surfaces $\beta$ and
$\delta_2$ in ZrAs$_2$ are quite complex, which makes it difficult
to distinguish the SdH oscillations of different FSs.
The calculated frequencies of $\delta_1$, $\delta_2$ and $\beta$
pockets are 135 T, 394 T and 551 T respectively, with magnetic
field along the $c$-axis. The frequency $F_1$=155 T may be
associated with a small Fermi surface hidden within $\delta_2$.
Combined with DFT calculations, we propose that $F_2$ is
associated with electron pocket $\delta_1$, $F_3$ is
associated with electron pocket $\delta_2$ and $F_4$ is associated
with hole pocket $\beta$ as shown in Fig. 4(a) - 4(c). However,
the oscillation frequency of small hole $\alpha$ pocket is not
observed in the SdH oscillations. The discrepancy between the SdH
experiments and the DFT calculations may be induced by the slight
variation in chemical potential of our sample. The chemical
potential is often influenced by the small changes in the lattice
parameters or impurity doping. In addition, the angle-dependent MR at 2 K
is shown in Fig. 4(g) with current along the $b$ axis and magnetic
field in the $ac$ plane. Besides, the obvious two-fold
symmetry of MR were obsrved and the ratio of $\rho_{xx}(14
T,(\varphi= 90^o))$ / $\rho_{xx}(14 T,(\varphi= 0^o))$ $\approx$ 3
also indicates a moderate electronic structure anisotropy
\cite{ref43}. The twofold FSs can acount for the two-fold symmetry
anisotropic MR \cite{ref48}. At 2 K and 14 T, the extremely large
unsaturated MR  reaches up to $1.9\times10^{4}$ \% with magnetic
field applied along the $c$ axis, which is significantly higher
than  that of $1.01\times10^{3}$ \% in previous report \cite{ref31}, indicating  higher quality of the single crystals.



In addition to the non-trivial Berry phase identified through quantum oscillation analysis, a compelling piece of evidence for the topological characteristics is the observation of chiral anomaly induced negative MR (NMR), which have been observed in Weyl semimetal NbAs and Dirac semimetal TaAs$_2$ and NbAs$_2$ \cite{ref9,ref10}. Fig. 5(a) shows the angle-dependent longitudinal MR at 2 K with magnetic field rotating from perpendicular to in parallel with current ($I // b$). In order to eliminate the influence of the Hall signal, the  $\rho_{xx}(B)$ were averaged by measuring the resitivity over positive and negative field directions. At $\theta= 0^o$, the MR is the largest. As $\theta$ decreases , the MR at 2 K decreases dramatically under magnetic field. When magnetic field is collinear with current ($\theta= 0^o$), the chiral anomaly plays a significant role in the transport. As shown in Fig. 5(b), when the magnetic field is parallel to current ($B // I$), a clear negative MR is observed. When the magnetic field increases, the MR becomes positive at low fields ($B < 5$ T) and negative at high fields below 50K, at $\theta= 90^o$. It should be noted that such a NMR was not in Ref. \cite{ref31}. Similar NMR induced by the chiral anomaly has widely observed in topological semimetals like NbAs$_2$ in Ref. \cite{ref9}. According to the theoretical works
and related experiments\cite{ref5,ref26,ref33}, when magnetic field
is collinear with electric field in the topological nodal line
semimetal, a chiral anomaly will appear remarkably which is similar to the
Weyl semimetal.  The chiral anomaly can account for the NMR when
the magnetic field is collinear with the electric field ($B //
I$), which indicates again the topological characteristic of ZrAs$_2$.

\begin{figure}[htbp]
\centering
  \includegraphics[width=0.48\textwidth]{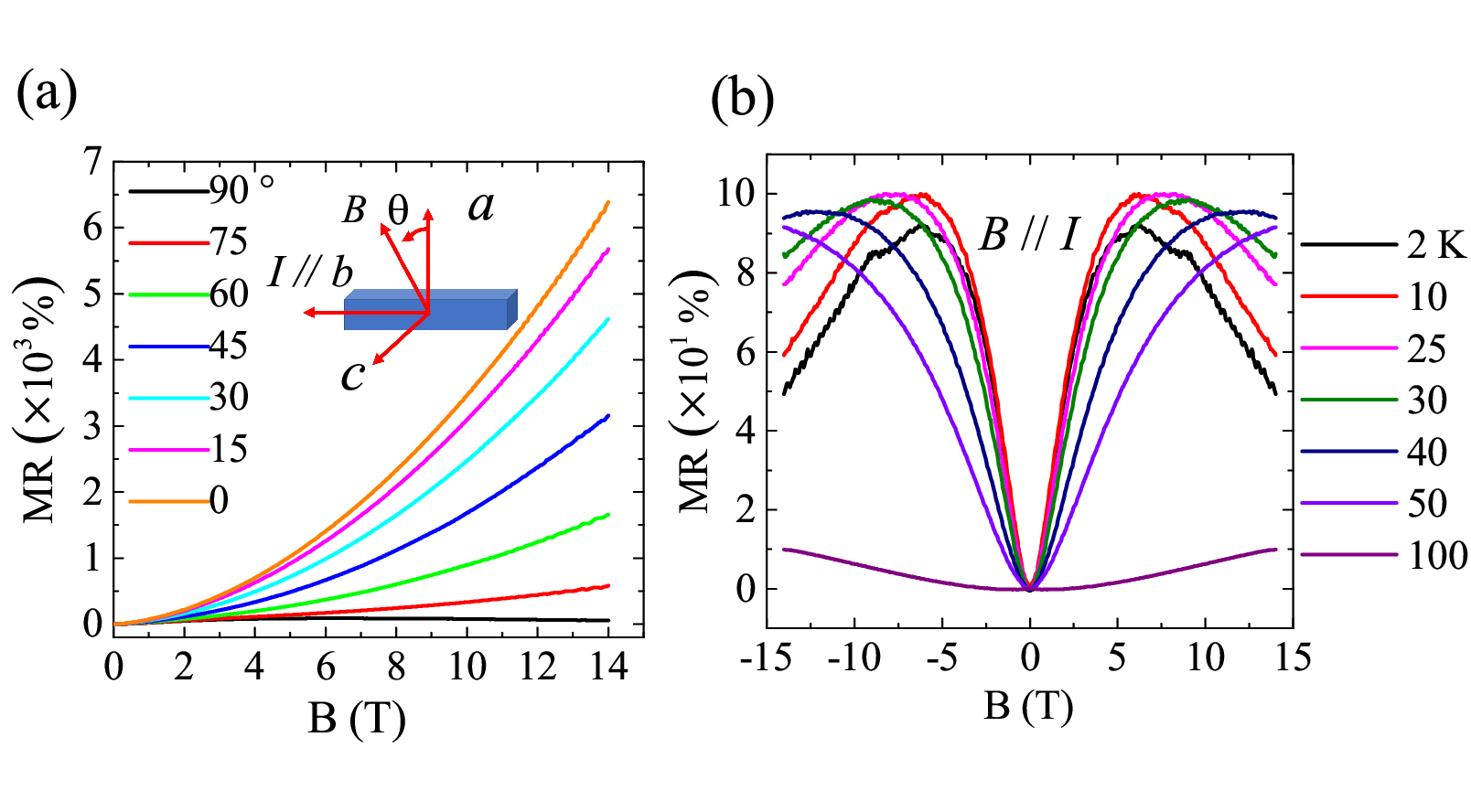}\\
\caption{(a) Angle-dependent MR as a function of magnetic field of ZrAs$_2$ by titling the magnetic fields from the a direction ($B$ $\bot$ $I$) to the $B$ // $I$ direction at 2 K. (b) Magnetic field-dependent MR at different temperatures with $B//I$ along the b-axis direction. } \label{6}
\end{figure}

Besides the chiral anomaly, there may be other mechanisms that can
also give rise to negative MR. Firstly, our ZrAs$_2$ sample is
high quality single crystalline which should not contain any
magnetic atoms, thus the influence of magnetic moments can be
excluded. Secondly, Anderson weak antilocalization (WAL) can also
cause negative MR at low fields \cite{ref49}. However, as shown in
Fig. 5(b), the negative MR in our case appears in high magnetic
field ($B > 5$ T), therefore the WAL effect could be excluded. Thirdly, the NMR can also be induced in the quantum limit, where the Fermi energy lies in the only the lowest Landau level \cite{ref50,ref51}. In our case, the quantum oscillations can be observed at high magnetic field as shown in Fig. 2(a), the influence of quantum limit can be
excluded. Finally, the current jetting effect can induce NMR
when inhomogeneous currents are injected into the sample. To
mitigate this, we positioned the current injection electrodes
across the sample to ensure a uniform current distribution
throughout the sample, thereby avoiding the influence of the
current jetting effect \cite{ref5}. To summarize, based on above discussions, and also considering theoretical calculations and analysis, we argued that the NMR in ZrAs$_2$ should be induced by the
chiral anomaly.

\section{CONCLUSIONS}

In summary, the detailed magnetotransport properties and electron
structures of topological nodal-line semimetal ZrAs$_2$ were
measured. The extremely large unsaturated MR is observed which
reaches up to $1.9\times10^{4}$ \% at 2 K and 14 T. The electron-
and hole-type charge carriers are almost compensated from the
analysis based on the two-band model, which may account for the
extremely large unsaturated MR at low temperature. The evident SdH
oscillations are observed and four frequencies $F_1 =$155 T , $F_2
=$178 T, $F_3 =$443 T and  $F_4 =$579 T are extracted from the SdH
oscillations. The nontrivial $\pi$ Berry phase and the chiral
anomaly induced NMR suggested its nontrivial topological
characteristic. Our study further confirmed the topological
characteristics of nodal-line semimetal ZrAs$_2$.
\section{Acknowledgments}

This work was supported by the National Science Foundation of
China (Grant No. 12174334; 12204410), the National Key \& Program
of the China (Grant No. 2019YFA0308602), the Innovation Program
for Quantum Science and Technology (Grant No. 2021ZD0302500), and
China Postdoctoral Science Foundation (Grant No. 2022M712788)

\end{document}